\documentclass[runningheads,a4paper]{llncs}
\usepackage{standalone}
\usepackage{hyperref}
\usepackage{pgf,pgfplotstable,tikz} 
\usetikzlibrary{positioning,calc,patterns}
\usepackage{pgfplots}
\pgfplotsset{compat=1.4}
\usepackage{tikz-3dplot}
\tdplotsetmaincoords{80}{30}
\tdplotsetrotatedcoords{0}{0}{0}%
\definecolor{c1}{rgb}{0.2,0.4,0.6} 
\definecolor{c2}{rgb}{1.0,0.0,0.6} 
\definecolor{c3}{rgb}{0.6,0.0,0.0} 
\usepackage{amssymb}
\usepackage{graphicx}
\usepackage{math}
\usepackage{url}
\usepackage{xcolor}
\usepackage{graphicx,float}

\urldef{\mailinria}\path|firstname.name@inria.fr|
\urldef{\mailiit}\path|saurabhk@iitk.ac.in|
\newcommand{\keywords}[1]{\par\addvspace\baselineskip
\noindent\keywordname\enspace\ignorespaces#1}

\pgfdeclarepatternformonly{my crosshatch dots}{\pgfqpoint{-1pt}{-1pt}}{\pgfqpoint{5pt}{5pt}}{\pgfqpoint{6pt}{6pt}}%
{
	\pgfpathcircle{\pgfqpoint{0pt}{0pt}}{.5pt}
	\pgfpathcircle{\pgfqpoint{3pt}{3pt}}{.5pt}
	\pgfusepath{fill}
}

\begin{document}

\mainmatter  

\title{VAST : The Virtual Acoustic \\ Space Traveler Dataset\vspace{-5mm}}

\titlerunning{VAST : The Virtual Acoustic Space Traveler Dataset}

%
%
\author{Cl\'{e}ment Gaultier$^{\star}$  \and Saurabh Kataria$^{\dagger}$ \and Antoine Deleforge$^{\star}$}
\authorrunning{Cl\'{e}ment Gaultier, Saurabh Kataria and Antoine Deleforge}

\institute{$^{\star}$Inria Rennes - Bretagne Atlantique, France (\mailinria)\\
$^{\dagger}$Indian Institute of Technology Kanpur, India (\mailiit)\vspace{-8mm}}

%
%

\toctitle{Lecture Notes in Computer Science}
\tocauthor{Authors' Instructions}
\maketitle

\begin{abstract}
This paper introduces a new paradigm for sound source localization referred to as virtual acoustic space traveling (VAST) and presents a first dataset designed for this purpose. Existing sound source localization methods are either based on an approximate physical model (physics-driven) or on a specific-purpose calibration set (data-driven). With VAST, the idea is to learn a mapping from audio features to desired audio properties using a massive dataset of simulated room impulse responses. This virtual dataset is designed to be maximally representative of the potential audio scenes that the considered system may be evolving in, while remaining reasonably compact. We show that virtually-learned mappings on this dataset generalize to real data, overcoming some intrinsic limitations of traditional binaural sound localization methods based on time differences of arrival.
\vspace{-5mm}
\end{abstract}

\section{Introduction}
\vspace{-3mm}
Human listeners have the stunning ability to understand complex auditory scenes using only two ears, \textit{i.e.}, with binaural hearing. Advanced tasks such as sound source direction and distance estimation or speech deciphering in multi-source, noisy and reverberant environments are performed daily by humans, while they are still a challenge for artificial (two-microphone) binaural systems. The main line of research in machine binaural source localization along 
the past decades has been to estimate the time-difference of arrival (TDOA) of the signal of interest at the two microphones. An estimated TDOA can be approximately mapped to the azimuth angle of a frontal source if the distance between microphones is known, assuming free-field\footnote{Free-field means that the sound propagates from the source to the microphones through a single direct path, without interfering objects or reverberations.} and far-field\footnote{Far-field means that the source is placed far enough (\textit{e.g.} $>1.8$ meters \cite{otani2009numerical}) from the receiver so that the effect of distance on recorded audio features is negligible.} conditions. Two important limits of these assumptions can be identified. First, they are both violated in most practical scenarios. In the example of an indoor binaural hearing robot, users are typically likely to engage interaction in both far- and near-fields and non-direct sound paths exist due to reflection and diffusion on walls, ceiling, floor, other objects in the room and the robot itself. Second, the intrinsic symmetries of a free-field/far-field binaural system restrict any geometrical estimation to that of a frontal azimuth angle. Hence, 3D source position (azimuth, elevation, distance) is out of reach in this scope, let alone additional properties such as source orientation, receiver position or room shape.

To overcome intrinsic limitations of TDOA, richer binaural features have been investigated. These include frequency-dependent phase and level differences \cite{viste2003use,deleforge2015co}, spectral notches \cite{raykar2005extracting,hornstein2006sound} or the direct-to-reverberant ratio \cite{lu2010binaural}. To overcome the free-field/far-field assumptions, advanced mapping techniques from these features to audio scene properties have been considered. These mapping techniques divide in two categories. The first one is \textit{physics-driven}, \textit{i.e.}, the mapping is inferred from an approximate sound propagation model such as the Woodworth's spherical head formula \cite{viste2003use}
or the full wave-propagation equation \cite{kitic2014hearing}. The second category of mapping is \textit{data-driven}. This approach is sometimes referred to as \textit{supervised sound source localization} \cite{talmon2011supervised}, or more generally \textit{acoustic space learning} \cite{deleforge2015acoustic}. These methods bypass the use of an explicit, approximate physical model by directly learning a mapping from audio features to audio properties using manually recorded training data \cite{talmon2011supervised,deleforge2015co}. They generally yield excellent results, but because obtaining sufficient training data is very time consuming, they only work for a specific room and setup and are hard to generalize in practice. Unlike artificial systems, human listeners benefit from years of adaptive auditory learning in a multitude of acoustic environments. While machine learning recently showed tremendous success in the field of speech recognition using massive amounts of annotated data, equivalent training sets do not exist for audio scene geometry estimation, with only a few specialized manually annotated ones \cite{deleforge2015acoustic,deleforge2015co}. Interestingly, a recent data-driven method \cite{parada2016single} used both real and simulated data to estimate room acoustic parameters and improve speech recognition performance, although it was not designed for sound localization.

We propose here a new paradigm that aims at making the best of physics-driven and data-driven approaches, referred to as \textit{virtual acoustic space traveling}. The idea is to use a physics-based room-acoustic simulator to generate arbitrary large datasets of room-impulse responses corresponding to various acoustic environments, adapted to the physical audio system considered. Such impulse responses can be easily convolved with natural sounds to generate a wide variety of audio scenes including \textit{cocktail-party} like scenarios. The obtained corpus can be used to learn a mapping from audio features to various audio scene properties using, \textit{e.g.}, deep learning or other non-linear regression methods \cite{deleforge2015high}. The \textit{virtually-learned} mapping can then be used to efficiently perform real-world auditory scene analysis tasks with the corresponding physical system. Inspired by the idea of an artificial system learning to hear by exploring virtual acoustic environments, we name this proposal the \textit{Virtual Acoustic Space Traveler} (VAST) project. We initiate it by publicly releasing a dedicated project page : \url{http://theVASTproject.inria.fr} and a first example of VAST dataset. This paper details the guidelines and methodology that were used in the process of building this training set. It then demonstrates that virtually-learned mappings can generalize to real-world test sets, overcoming intrinsic limitations of TDOA-based sound source localization methods.

\vspace{-3mm}
\section{Dataset Design}
\vspace{-3mm}
\subsection{General Principles}
\vspace{-3mm}
The space of all possible acoustic scenes is vast. Therefore, some trade-offs between the size and the representativity of the dataset must be made when building a training corpus for audio scene geometry estimation. During the process of designing the dataset, we imposed on ourselves the following guidelines:
\vspace{-3mm}
\begin{itemize}
 \item[$\bullet$] The dataset should consist of room impulse responses (RIR). This is a more generic representation than, \textit{e.g.}, specific audio features or audio scenes involving specific sounds. Each RIR should be annotated by all the source, receiver and room properties defining it.
 \item[$\bullet$]  Virtual acoustic space traveling aims at building a dataset for a \textbf{specific audio system} in a variety of environments. Following this idea, some intrinsic properties of the receiver such as its distance to the ground and its head-related transfer functions are kept fixed throughout the simulations. For this first dataset, called \textit{VAST\_KEMAR\_0}, we chose the emblematic KEMAR acoustic dummy-head, whose measured HRTFs are publicly available. It was placed at 1.70 from the ground, the average human's height.
 \item[$\bullet$]  We are interested in modeling acoustic environments which are typically encountered in an office building, a university, a hotel or a modern habitation. Acoustics of the type encountered in a cathedral, a massive hangar, a recording studio or outdoor are deliberately left aside here. Surface materials and diffusion profiles are chosen accordingly.
 \item[$\bullet$]  To make the dataset easily manipulable on a simple laptop, we aimed at keeping its total size under 10 GigaBytes. To handle datasets of larger order of magnitudes would require users to have access to specific hardware and software which is not desired here. \textit{VAST\_KEMAR\_0} measures 6.4 GB.
\end{itemize}
\vspace{-6mm}
\vspace{-3mm}
\subsection{Room Simulation and Data Generation}
\vspace{-3mm}
The efficient C++/MATLAB ``shoebox'' 3D acoustic room simulator ROOMSIM developed by Schimmel et al. is selected for simulations \cite{schimmel2009fast}. This software takes as input a room dimension (width, depth and height), a source and receiver position and orientation, a receiver's head-related-transfer function (HRTF) model, and frequency-dependent absorption and diffusion coefficients for each surface. It outputs a corresponding pair of room impulse responses (RIR) at each ear of the binaural receiver. Specular reflections are modeled using the image-source method \cite{allen1979image}, while diffusion is modeled using the so-called \textit{rain-diffusion} algorithm. In the latter, sound rays uniformly sampled on the sphere are sent from the emitter and bounced on the walls according to specular laws, taking into account surface absorption. At each impact, each ray is also randomly bounced towards the receiver with a specified probability (the frequency-dependent \textit{diffusion coefficient} of the surface). The total received energy at each frequency is then aggregated using histograms. This model was notably showed to realistically account for sound scattering due to the presence of objects, by comparing simulated RIRs with measured ones in \cite{wabnitz2010room}. The study \cite{kataria2016hearing} suggests that such diffusion effects play an important role in sound source localization. \textit{VAST\_KEMAR\_0} contains over $110,000$ RIR, which required about 700 CPU-hours of computation. This was done using a massively parallelized implementation on a large computer grid.

\vspace{-8mm}
\begin{figure}
\centering
    \begin{tabular}{cc}
	        \includegraphics[height=4.0cm]{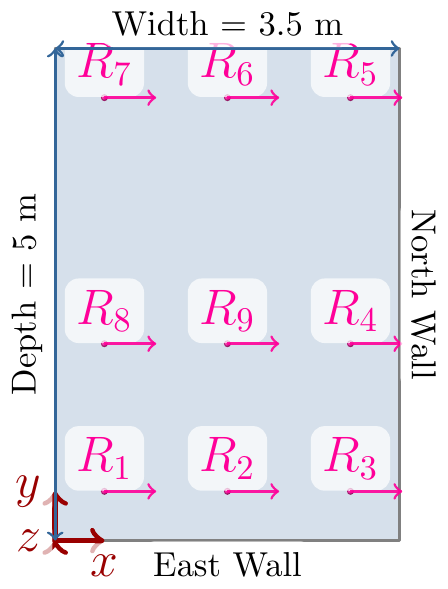} & 
		\includegraphics[height=4.8cm]{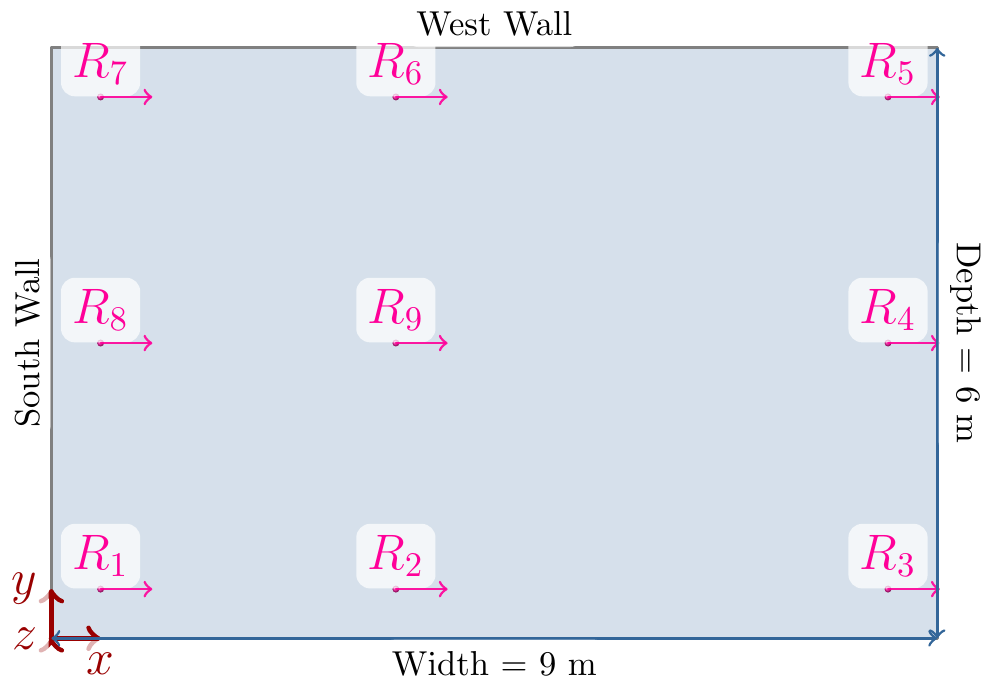} \\
		(a) Small Room (Height = 2.5m)& (b) Large Room (Height = 3.5m)
    \end{tabular}
    \vspace*{-2mm}    
    \caption{Top views of training rooms with receiver positions and orientations.}
    \vspace*{-5mm}
    \label{fig:TopView}
\end{figure}
\begin{figure}
	\vspace*{-5mm}
		\centering
                \includegraphics[width=0.8\textwidth]{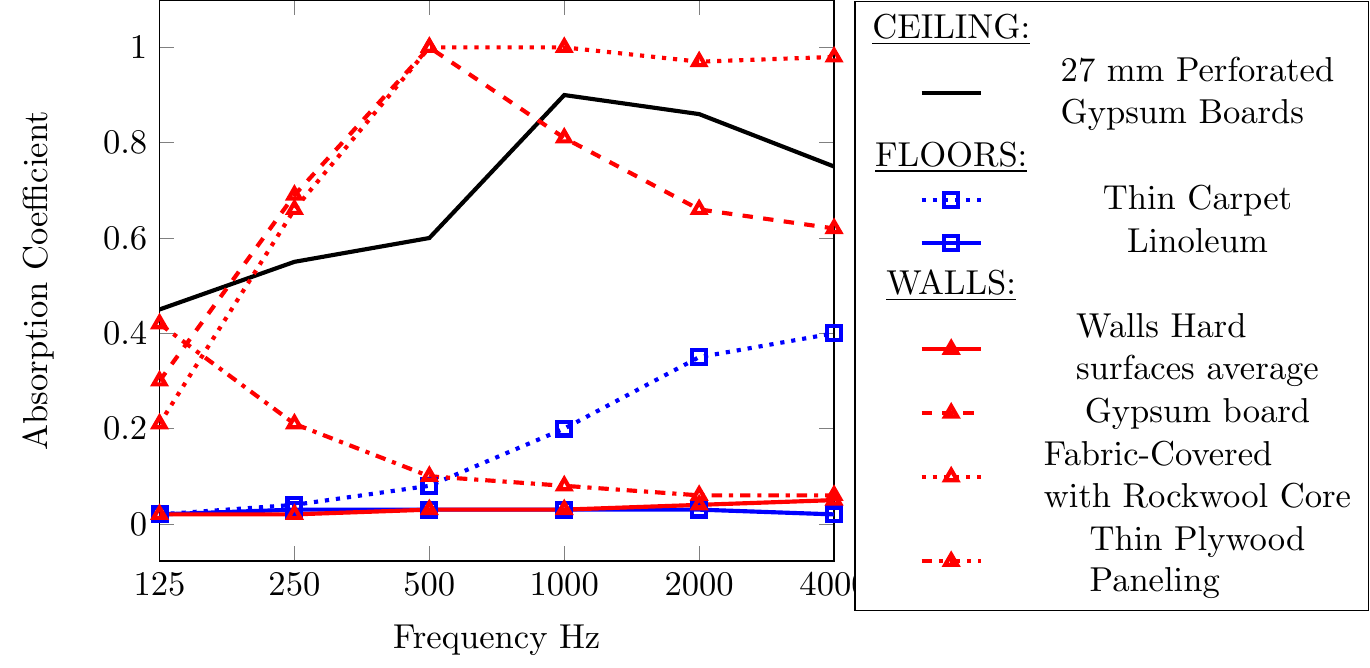}
	\vspace*{-4mm}		
		\caption{Absorption Profiles}
	\label{fig:AbsorptionProfiles}
	\vspace*{-5mm}
\end{figure}
\vspace{-6mm}
\subsection{Room Properties: Size and Surfaces}
\vspace{-3mm}
An obvious choice to generate virtual rooms with maximal variability would be to draw a random room size and random frequency-dependent absorption and diffusion profiles of surfaces for each generated RIR. This approach however, has several drawbacks. First, it makes impossible the generation of realistic audio scenes containing several sources, for which the receiver position and the room must be fixed. Second, the space of possible rooms is so vast that reliably sampling it at random is unrealistic. Third, changing source, receiver and room parameters all at the same time prevents from getting insights on the individual influence of these parameters. On the other hand, sampling all combinations of parameters in an exhaustive way quickly leads to enormous data size. As a trade-off, we designed 16 realistic rooms representative of typical reverberation time ($RT_{60}$) and surface absorption profiles encountered in modern buildings. Two room sizes were considered: a small one corresponding to a typical office or bed room (Fig.~\ref{fig:TopView}(a)), and a larger one corresponding to a lecture or entrance hall (Fig.~\ref{fig:TopView}(b)). For each room, floor, ceiling and wall materials which are representative in terms of absorption profile and are commonly encountered in nowadays buildings were chosen from~\cite{vorlander2007auralization}. The graph on Fig.~\ref{fig:AbsorptionProfiles} displays the absorption profiles of the selected materials, namely, 4 for the walls, 2 for the floor and 1 for the ceiling. The gypsum board material chosen for the ceiling was kept fixed throughout the dataset, as it represents well typical ceiling absorption profiles~\cite{vorlander2007auralization}. ``Walls hard surface average'' is in fact an average profile over many surfaces such as brick or plaster~\cite{vorlander2007auralization}. Combining all possible floors, walls and room sizes yielded the 16 rooms listed in Table~\ref{tab:RoomDescript}.

Importantly, typical rooms also contain furniture and other objects responsible for random sound scattering effects, \textit{i.e.}, diffusion. Following the acoustic study in \cite{faiz2012measurement}, a unique frequency-dependent diffusion profile was used for all surfaces. The chosen profile is the average of the 8 configurations measured in~\cite{faiz2012measurement}, corresponding to varying numbers of chairs, table, computers and people in a room. Both absorption and diffusion profiles are piecewise-linearly interpolated from 8 Octave bands from 125 Hz to 4 kHz.

\begin{table}[t]
	\centering
	\caption{Description of simulated training rooms in VAST}	
	\label{tab:RoomDescript}
	\resizebox{\textwidth}{!}{%
		\begin{tabular}{|c|l|l|l|c|c|c|}
			\hline
			\textbf{\begin{tabular}[c]{@{}c@{}}Room\\ Number\end{tabular}} & \multicolumn{1}{c|}{\textbf{Floor}} & \multicolumn{1}{c|}{\textbf{Ceiling}} & \multicolumn{1}{c|}{\textbf{Walls}}     & \textbf{\begin{tabular}[c]{@{}c@{}}Width\\ {[}m{]}\end{tabular}} & \textbf{\begin{tabular}[c]{@{}c@{}}Depth\\ {[}m{]}\end{tabular}} & \textbf{\begin{tabular}[c]{@{}c@{}}Height\\ {[}m{]}\end{tabular}} \\ \hline		
			1                    & Thin Carpet                         & Perforated 27 mm gypsum board         & Walls Hard Surfaces Average             & 9              & 6              & 3.5             \\ \hline
			2                    & Thin Carpet                         & Perforated 27 mm gypsum board         & Gypsum Board with Mineral Filling       & 9              & 6              & 3.5             \\ \hline
			3                    & Thin Carpet                         & Perforated 27 mm gypsum board         & Fabric-Covered Panel with Rockwool Core & 9              & 6              & 3.5             \\ \hline
			4                    & Thin Carpet                         & Perforated 27 mm gypsum board         & Thin Plywood Paneling                  & 9              & 6              & 3.5             \\ \hline
			5                    & Linoleum                            & Perforated 27 mm gypsum board         & Walls Hard Surfaces Average             & 9              & 6              & 3.5             \\ \hline
			6                    & Linoleum                            & Perforated 27 mm gypsum board         & Gypsum Board with Mineral Filling       & 9              & 6              & 3.5             \\ \hline
			7                    & Linoleum                            & Perforated 27 mm gypsum board         & Fabric-Covered Panel with Rockwool Core & 9              & 6              & 3.5             \\ \hline
			8                    & Linoleum                            & Perforated 27 mm gypsum board         & Thin Plywood Paneling                  & 9              & 6              & 3.5             \\ \hline
			9                    & Thin Carpet                         & Perforated 27 mm gypsum board         & Walls Hard Surfaces Average             & 3.5            & 5              & 2.5             \\ \hline
			10                   & Thin Carpet                         & Perforated 27 mm gypsum board         & Gypsum Board with Mineral Filling       & 3.5            & 5              & 2.5             \\ \hline
			11                   & Thin Carpet                         & Perforated 27 mm gypsum board         & Fabric-Covered Panel with Rockwool Core & 3.5            & 5              & 2.5             \\ \hline
			12                   & Thin Carpet                         & Perforated 27 mm gypsum board         & Thin Plywood Paneling                  & 3.5            & 5              & 2.5             \\ \hline
			13                   & Linoleum                            & Perforated 27 mm gypsum board         & Walls Hard Surfaces Average             & 3.5            & 5              & 2.5             \\ \hline
			14                   & Linoleum                            & Perforated 27 mm gypsum board         & Gypsum Board with Mineral Filling       & 3.5            & 5              & 2.5             \\ \hline
			15                   & Linoleum                            & Perforated 27 mm gypsum board         & Fabric-Covered Panel with Rockwool Core & 3.5            & 5              & 2.5             \\ \hline
			16                   & Linoleum                            & Perforated 27 mm gypsum board         & Thin Plywood Paneling                  & 3.5            & 5              & 2.5             \\ \hline
			0                    & \multicolumn{6}{l|}{Anechoic room} \\ \hline
			\end{tabular}
	}
	\vspace*{-5mm}
\end{table}

\vspace{-3mm}
\subsection{Reverberation Time}
\begin{figure}
\centering
    \begin{tabular}{cc}
	        \includegraphics[height=4.0cm]{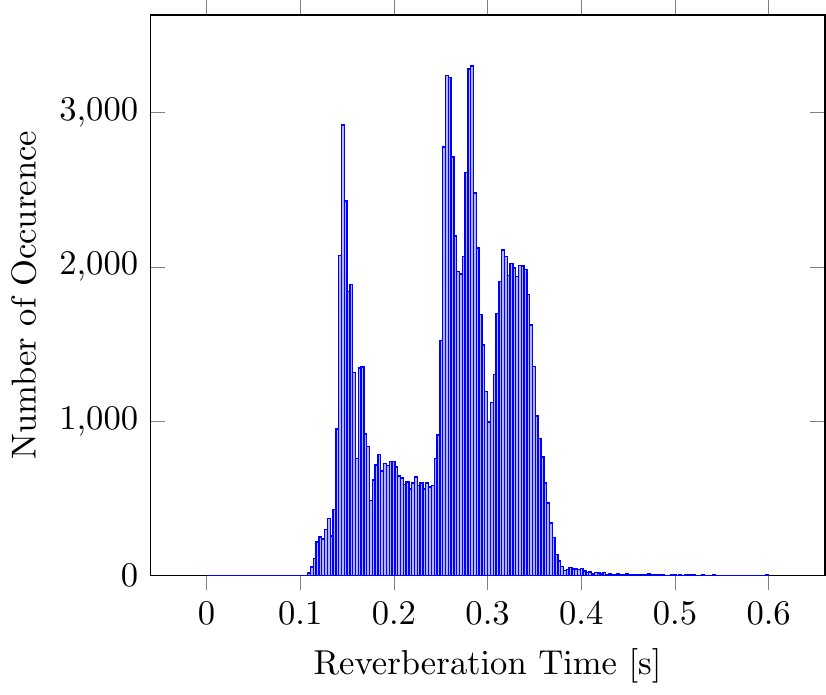} & 
	        \includegraphics[height=4.0cm]{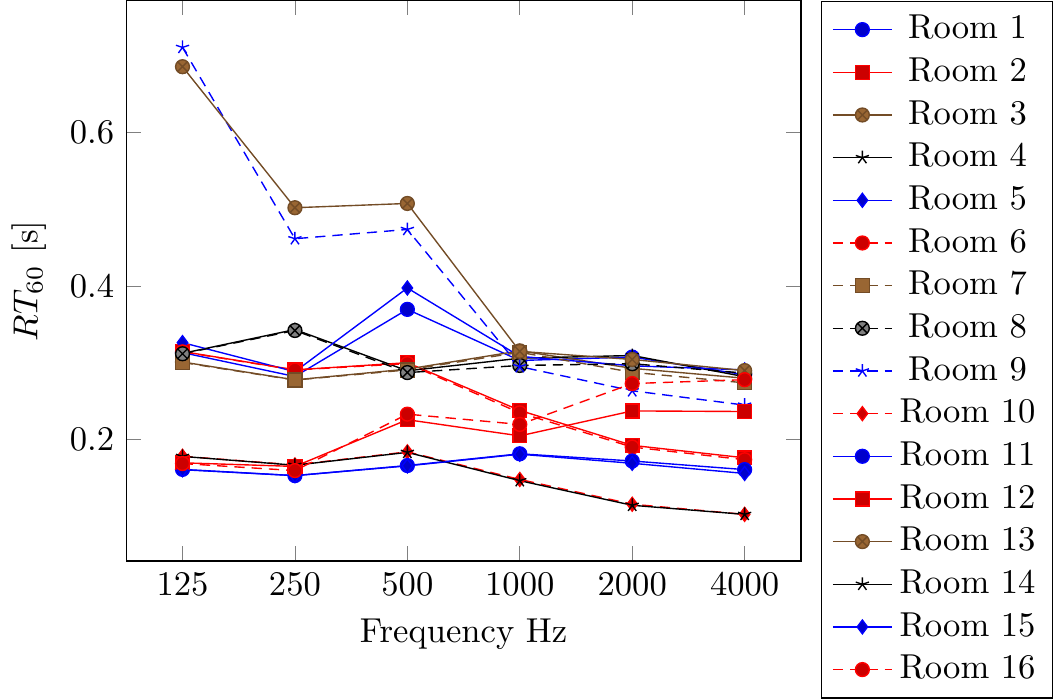} \\ 
		(a) Global $RT_{60}$ distribution & (b) Octave bands $RT_{60}$
    \end{tabular}
    \vspace*{-3mm}    
    \caption{Reverberation Time.}
    \vspace*{-6mm}
    \label{fig:RT60}
\end{figure}

\vspace{-3mm}
A common acoustic descriptor for rooms is the reverberation time ($RT_{60}$). Figure \ref{fig:RT60}(a) displays the estimated $RT_{60}$ distribution across the VAST Training Dataset. Fig.~\ref{fig:RT60}(b) shows the $RT_{60}$ for each room by octave band. $RT_{60}$'s were estimated from the room impulse responses following the recommendations in~\cite{schroeder1965new}. From these estimations, we decided to crop the room impulse responses provided in the datasets above the $RT_{60}$, with a 30 ms margin. This technique allows to shrink the dataset while keeping data points of interest and discarding the rest. To further complies with memory limitations, we chose to encode the room impulse response samples with single floats (16 bit). As can be seen in Fig.~\ref{fig:RT60} the 16 chosen rooms present a quite good variability in terms of reverberation times in the range 100ms-400ms. Larger $RT_{60}$ of the order of 1 second could be obtain by using highly reflective materials on all surfaces, creating an echo chamber. However, this rarely occurs in realistic buildings. 
\vspace{-6mm}
\vspace{-7mm}	
\subsection{Source and Receiver Positions}
\vspace{-3mm}	
A relatively poorly-studied though important effect in sound source localization is the influence of the receiver's position in the room, especially its distance to the nearest surface. In order to accurately capture this effect, 9 receiver positions are used for each of the 16 rooms, while the height of the receiver is fixed at 1.7 m. Figure~\ref{fig:TopView} shows top views of the rooms with receiver positions. Positions from $R_1$ to $R_8$ are set 50 cm from the nearest wall(s) whereas $R_9$ is approximately placed in the middle of the room. Perfectly symmetrical configurations are avoided to make the dataset as generic as possible, without singularities. The receiver is always facing the north wall as a convention. For each of the 9 receiver positions, sources are placed on spherical grids centered on the receiver. Each sphere consists of regularly-spaced elevation lines each containing sources at regularly-spaced azimuths, with a spacing of $9^\circ$. The equator elevation line and the first azimuth angle of each line are randomly offset by $-4.5^\circ$ to $+4.5^\circ$ in order to obtain a dense sphere sampling throughout the dataset. Six spherical grid radii are considered, yielding source distances of 1, 1.5, 2, 3, 4 and 6 meters. Sources falling outside of the room or less than 20cm from a surface are removed.

\vspace{-3mm}
\subsection{Test Sets}
\label{subset:test_set}
\vspace{-3mm}
To test the generalizability of mappings learned on the \textit{VAST\_KEMAR\_0} dataset, we built four simulated test sets differing from the training dataset on various levels. 
A first challenge is to test robustness to random positioning, since the training set is built with regular spherical source grids and fixed listener positions. Hence, the 4 testing sets contain completely random source and receiver positions in the room. Only the receiver's height is fixed to 1.7m, and both receiver and source are set within a 20 cm safety margin within the room boundaries. Test sets 2 and 4 feature random receiver orientation (yaw angle), as opposed to the receiver facing north in the training set. Test 1 and 2 contain 1,000 binaural RIRs (BRIRs) for each of the 16 rooms of Table \ref{tab:RoomDescript}. Finally, test sets 3 and 4 contain 10,000 BRIRs, each corresponding to a random room size (walls from $3m\times2m$ to $10m\times4m$) and random absorption properties of walls and floor picked from Fig.~\ref{fig:AbsorptionProfiles}. Different surfaces for all 4 walls are allowed.

In addition to these simulated test sets, three binaural RIR datasets recorded with the KEMAR dummy head in real rooms have been selected, as listed below:
\vspace{-6mm}
\begin{itemize}
 \item[$\bullet$] \textbf{Auditorium 3 \cite{ma2015machine}} was recorded at TU Berlin in 2014 in a trapezium-shaped lecture room of dimensions 9.3m $\times$ 9m and $RT_{60}\approx0.7$s. 3 individual sources placed 1.5m from the receiver at different azimuth and $0^\circ$ elevation were recorded. For each source, one pair of binaural RIR is recorded for each receivers' head yaw angle from $-90^\circ$ to $+90^\circ$, with $1^\circ$ steps.
 \item[$\bullet$] \textbf{Spirit \cite{ma2015machine}} was recorded at TU Berlin in 2014 in a small rectangular office room of size 4.3m $\times$ 5m, $RT_{60}\approx0.5$s, containing various objects, surfaces and furniture near the receiver. The protocol is the same as Auditorium 3 except sources are placed 2m from the receiver.
 \item[$\bullet$] \textbf{Classroom \cite{shinn2005localizing}} was recorded at Boston University in 2005 in a 5m $\times$ 9m $\times$ 3.5m carpeted classroom with 3 concrete walls and one sound-absorptive wall ($RT_{60}=565$ms). The receiver is placed in 4 locations of the room including 3 with at least one nearby wall. 
\end{itemize}
\vspace{-3mm}
Note that the KEMAR HRTF measurements used to simulate the VAST dataset was recorded by yet another team, in MIT's anechoic chamber in 1994, as described in \cite{gardner1995hrtf}.

\vspace{-3mm}
\section{Virtually Supervised Sound Source Localization}
\vspace{-3mm}
For all experiments in this section, all training and test sets used are reduced to contain only frontal sources (azimuth in $[-90^\circ,+90^\circ]$) with elevation in $[-45^\circ$, $+45^\circ]$ and distances between 1 and 3 meters.
As mentioned in the introduction, sound source localization consists in two steps: calculating auditory features from binaural signals followed by mapping these features to a source position. Robustly estimating features can be difficult when dealing with additive noise, sources with sparse spectra such as speech or music, and source mixtures. We leave this problematic aside in this paper, and focus on mapping clean features to source positions. Hence, we use \textit{ideal} features directly calculated from the clean room impulse responses in all experiments.

\begin{figure}
	\vspace*{-3mm}
		\centering
		\includegraphics[width=\textwidth]{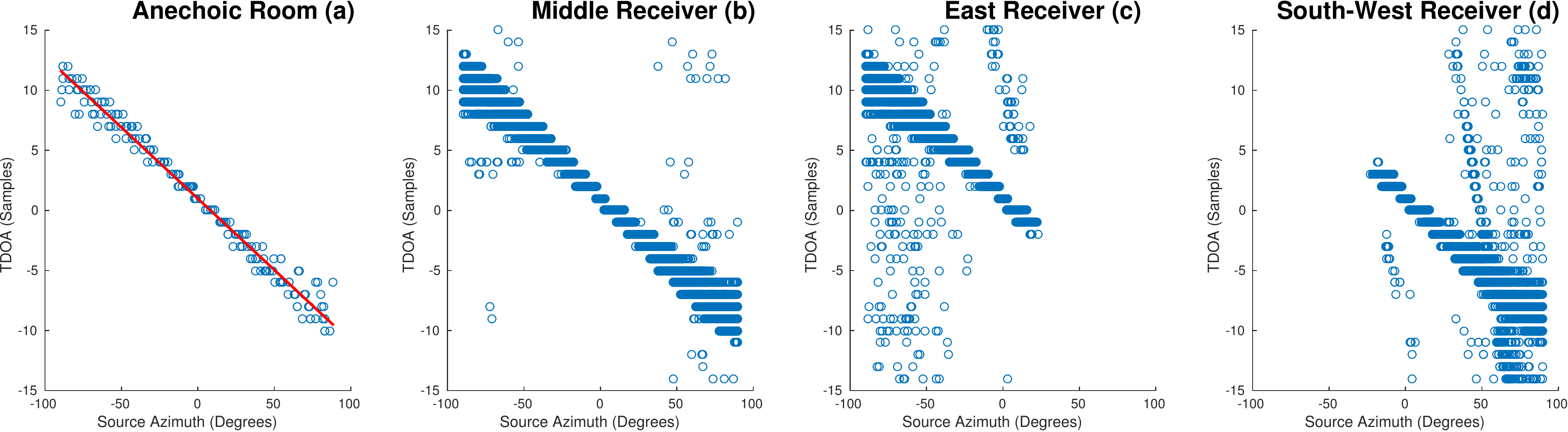}
		\vspace{-6mm}
		\caption{TDOA as a function of source azimuth in various settings}
	\label{fig:TDOAaz}
\end{figure}

We first make an experiment to put forward some intrinsic limitations of TDOA-based azimuth estimation. Fig.~\ref{fig:TDOAaz} plots TDOAs against the source's azimuth angle for different subsets of VAST. TDOAs (in samples) were computed as the integer delay in $[-15,+15]$ maximizing the correlation between the first 500 samples of the left and the right impulse responses. As can be seen in Fig.~\ref{fig:TDOAaz}(a), a near-linear relationship between frontal azimuth and TDOA exists in the anechoic case, regardless of the elevation. This matches previously observed results in binaural sound localization \cite{viste2003use,sanchez2012online,deleforge2015co}. When the receiver is placed in the middle of the 16 reverberant rooms, (Fig.~\ref{fig:TDOAaz}(b)), some outliers appear due to reflections. This effect is dramatically increased when the receiver is placed 50 centimeters from a wall (Fig.~\ref{fig:TDOAaz}(c) and \ref{fig:TDOAaz}(d)), where stronger early reflections are present. This suggests that the TDOA, even when ideally estimated, is not adapted to binaural sound source localization in realistic indoor environments.

\begin{table}[htb]
	\centering
	\caption{Azimuth absolute estimation errors in degrees with 3 different methods, showed in the form $\textit{avg}\pm\textit{std}  (\textit{out}\%)$, where $\textit{avg}$ and $\textit{std}$ denote the mean and standard deviation of \textit{inlying} absolute errors ($<30^\circ$) while out denotes the percentage of outliers.}
	\label{tab:results_az}
	\vspace{-3mm}
	\resizebox{\textwidth}{!}{%
	\begin{tabular}{l|c|c|c}
	\textbf{Test data $\downarrow$} & TDOA & GLLiM (Anech. train.)& GLLiM (VAST train.) \\
	\hline
	\textbf{VAST Testing Set 1}                   & $5.49\pm4.6 (5.6\%)$ & $8.63\pm7.6 (12\%) $ & $4.38\pm4.9 (1.8\%)$  \\
	\textbf{VAST Testing Set 2}                   & $5.37\pm4.4 (6.0\%)$ & $8.09\pm7.5 (12\%) $ & $4.32\pm4.7 (1.6\%) $  \\
	\textbf{VAST Testing Set 3}                   & $5.21\pm4.5 (4.6\%)$ & $8.46\pm7.5 (5.2\%)$ & $4.23\pm4.4 (1.8\%)$  \\
	\textbf{VAST Testing Set 4}                   & $5.14\pm4.4 (3.3\%)$ & $8.21\pm7.2 (4.8\%)$ & $4.25\pm4.4 (0.6\%)$  \\
	\textbf{Auditorium 3 \cite{ma2015machine}}    & $7.02\pm4.7 (1.4\%)$ & $8.01\pm7.0 (5.9\%)$ & $5.03\pm4.5 (0.0\%)$  \\	
	\textbf{Spirit \cite{ma2015machine}}          & $5.19\pm3.4 (0.0\%)$ & $12.2\pm8.3 (15\%) $ & $4.50\pm5.6 (0.4\%)$  \\
	\textbf{Classroom \cite{shinn2005localizing}} & $5.71\pm3.7 (3.7\%)$ & $9.47\pm7.3 (5.2\%)$ & $6.50\pm5.9 (0.0\%)$  \\	
	\end{tabular}	
	}
	\vspace{-3mm}
\end{table}

\begin{table}[htb]
	\centering
	\caption{Elevation and distance absolute estimation errors obtained with GLLiM trained on VAST. Outliers correspond to errors larger than $15^\circ$ or 1m.}
	\label{tab:results_extra}
	\vspace{-3mm}
	\begin{tabular}{l|c|c}
	\textbf{Test data $\downarrow$} & Elevation (${}^\circ$) & Distance (m) \\
	\hline
	\textbf{VAST Testing Set 1}                   & $5.91\pm4.1 (23\%) $ & $0.43\pm0.3 (19\%)$ \\
	\textbf{VAST Testing Set 2}                   & $6.05\pm4.2 (27\%) $ & $0.44\pm0.3 (20\%)$  \\
	\textbf{VAST Testing Set 3}                   & $6.05\pm4.1 (27\%) $ & $0.43\pm0.3 (21\%)$  \\
	\textbf{VAST Testing Set 4}                   & $6.03\pm4.2 (26\%) $ & $0.44\pm0.3 (21\%)$  \\
	\textbf{Auditorium 3 \cite{ma2015machine}}    & $7.92\pm4.4 (44\%) $ & $0.45\pm0.3 (23\%)$  \\	
	\textbf{Spirit \cite{ma2015machine}}          & $7.44\pm4.3 (30\%) $ & $0.52\pm0.3 (25\%)$   \\
	\textbf{Classroom \cite{shinn2005localizing}} & $8.40\pm4.1 (45\%) $ & $0.41\pm0.3 (6.5\%)$  \\	
	\end{tabular}	
	\vspace{-3mm}
\end{table}

We then compare azimuth estimation errors obtained with the TDOA-based method described above, a learning-based method trained on anechoic HRTF measurements (Room 0), and a learning-based method trained on VAST, using the 4 simulated and 3 real test sets described in Section \ref{subset:test_set}. TDOAs were mapped to azimuth values using the affine regression coefficients corresponding to the red line in Fig.~\ref{fig:TDOAaz}(a). The chosen learning-based sound source localization method is the one described in \cite{deleforge2015co}. It uses Gaussian Locally Linear Regression (GLLiM, \cite{deleforge2015high}) to map high-dimensional feature vectors containing frequency-dependent interaural level and phase differences from 0 to 8000 Hz to low-dimensional source positions. In our case, the GLLiM model with $K$ locally-linear components was trained on $N$ interaural feature vectors of dimension $D=1537$ associated to 3-dimensional source positions in spherical coordinate (azimuth, elevation and distance). $K=8$ components were used for the anechoic training set ($N=181$) and $K=100$ for the (reduced) VAST dataset ($N\approx41,000$). All 3 methods showed comparably low testing computational times, in the order of 10ms for 1 second of input signal.
Table \ref{tab:results_az} summarizes obtained azimuth estimation errors. As can be seen, the learning method trained on VAST outperforms the two others on all datasets, with significantly less outliers and a globally reduced average error of inliers. This is encouraging considering the variety of testing data used.
In addition, Table \ref{tab:results_extra} shows that GLLiM trained on VAST is capable of approximately estimating the elevation and distance of the source, which is known to be particularly difficult from binaural data. While elevation estimation on real data remains a challenge, results obtained on simulated sets are promising.

\vspace{-4mm}
\section{Conclusion}
\vspace{-3mm}
We introduced the new concept of virtual acoustic space traveling and released a first dataset dedicated to it. A methodology to efficiently design such a dataset was provided, making extensions and improvements of the current version easily implementable in the future. Results show that a learning-based sound source localization method trained on this dataset yields better localization results than when trained on anechoic HRTF measurements, and performs better than a TDOA-based approach in azimuth estimation while being able to estimate source elevation and distance. To the best of the authors' knowledge, this is the first time a sound localization method trained on simulated data is successfully used on real data, validating the new concept of virtual acoustic space traveling. The learning approach could still be significantly improved by considering other auditory features, by better adapting the mapping technique to spherical coordinates and by annotating training data with further acoustic information. Other learning methods such as deep neural networks may also be investigated.

\vspace{-4mm}
\bibliographystyle{splncs03}
\bibliography{refs_lva2017_gaultier}

\end{document}